\documentclass[10pt]{article}

\usepackage{amsmath}
\usepackage{amssymb}

\usepackage{graphicx}

\usepackage{cite}

\usepackage{axodraw}

\usepackage{color} 

\usepackage{subfig}

\makeatletter
\renewcommand{\@biblabel}[1]{\quad#1.}
\makeatother

\date{}

\begin{document}

\begin{flushleft}
{\Large
\textbf{The importance of being discrete in sex}
}
\\
Author1$^{1}$, 
\\
\bf{1} Renato Vieira dos Santos Departamento de F\'{i}sica, Instituto de Ci\^{e}ncias Exatas, Universidade Federal de Minas Gerais, CP 702,
CEP 30161-970, Belo Horizonte, Minas Gerais, Brasil.
\\
$\ast$ E-mail: econofisico@gmail.com
\end{flushleft}

\begin{abstract}

The puzzle associated with the cost of sex, an old problem of evolutionary biology, is discussed here from the point of view of
nonequilibrium statistical mechanics. The results suggest, in a simplified model, that the prevalence of sexual species in nature can be a
natural and necessary consequence of the discrete character of the nonlinear interactions between couples and their pathogens/parasites.
Mapped into a field theory, the stochastic processes performed by the species are described by continuous fields in space and time. The way
that the model's parameters scale with subsequent iterations of the renormalization group gives us information about the stationary
emergent properties of the complex interacting systems modeled. We see that the combination of one aspect of the Red Queen theory with the
stochastic processes theory, including spatiotemporal interactions, provides interesting insights into this old Darwinian dilemma.
\end{abstract}


\section{Introduction}
\label{intro}

Sex is an evolutionary puzzle. In several ways, sexual reproduction is less efficient when compared with the asexual method.
All offspring produced by asexual individuals will be able to reproduce, whereas sexual beings need to spend energy on creating males and
females that do not reproduce separately. Hence the resources spent on producing sons are a cost of sexual reproduction and asexual species
economize on males. John Maynard Smith \cite{smith1993theory} summarized this argument as follows: 

\begin{quotation}

\textit{``Suppose a population consists of a mixture of sexual and parthenogenetic females, the former producing equal numbers of male
and (sexual) female offspring, and the latter only parthenogenetic females like themselves. If the two kinds of female lay equal numbers of
eggs, and if survival probabilities are equal, then the parthenogenetic type will have a twofold selective advantage, and will increase in
frequency very rapidly. Sexual reproduction means that a female wastes half her energy producing males.''}

\end{quotation}

He also noted that a sexual individual uses only half of its genetic material on its descendents, while an asexual individual
uses all his sexless genes. That is, in the evolutionary race where passing on genes to the next generation is one of the greatest goals,
sexual organisms starts with a disadvantage of almost $50\%,$ which is known as the \emph{cost of meiosis}. There is also the risk of
infection by sexually transmitted diseases. In addition to these disadvantages, and perhaps more crucial, we should mention the cost
of having to find a mate.
If insects are excluded, approximately one-third of animal species are hermaphrodites
\cite{jarne2006animals}. Hermaphroditism is even more widespread in plants. The difficulty of finding mates is widely implicated in the
evolution of hermaphroditism, so its widespread occurrence suggests that sexual organisms incur significant costs to locate mates
\cite{eppley2008moving}. Sex, therefore, seems to be a luxury that should not exist. Consequently, many works about its evolution look
for it's compensatory benefits.

Since sexual reproduction exists, biologists try to find out what great benefit it brings to living beings. Maynard Smith argued that sex
could only have evolved if this mysterious benefit at least outweighed the great cost of meiosis. But what, after all, could this benefit
be? To answer this question, an audacious theory about the origin and perpetuation of sex was proposed in \cite{hamilton1982heritable}.
According to this work, the parasites \emph{are everywhere and will always seek, by their nature, to explore their hosts.}
As the generation time of parasites is many times smaller than that of hosts, and their evolution rates therefore many times higher, the
only way out for the hosts is to produce offspring with greater genetic variability through sexual reproduction.
Therefore, competition with parasites that develop very fast genetically, favors sexual reproduction, which enables a more
efficient genetic evolution.

The world in which this model is inserted became known as the Red Queen's world, a name given in \cite{valen73} in reference to a passage in
the fable Alice in the mirrors \cite{carroll2000alice}. In this passage, Alice flees the army (of cards) of the Red Queen, but can not
distance herself from her pursuers. The Red Queen then says: ``Now, here, you see, it takes all the running you can do, to keep in the same
place''. Alice would be caught only if she stopped running. Things have to change to remain the same.

According to \cite{hamilton1982heritable}, an arms race has been underway between hosts and parasites since life appeared on Earth.
The parasites are always breaking the defensive barriers imposed by the host's genotype, while the host, with the help of sex, continually
creates new defenses. In the absence of sex, the hosts would remain essentially the same, while the parasites would accumulate adaptations
that would enable them to break all the defensive systems of the former. Sooner or later, the hosts would be virtually devoured from the
inside out. To escape the parasite army besieging them, the only remaining option is to just keep running. The co-evolutive cycle of
parasites and hosts reflects this eternal pursuit.

The aim of this paper is to investigate the prevalence of sexual reproduction observed in nature through simple models of
reaction-diffusion inspired by the Red Queen theory, but not fully equivalent to it. The role of the parasite may be replaced by any
pathogen which diffuses through space and fatally harms the species. There is also no need for any aspect related to genetics.

We take into account the discrete nature of the species interactions with itself and with its pathogens /parasites. We know that if we want
information about the emergent aggregate macroscopic behavior of complex systems, we'll need to consider the corpuscular character of
interacting species \cite{shnerb_discrete}. We will achieve this goal by using the dynamic renormalization group (DRG) theory to
obtain the renormalization group (RG) flow in the parameter space. Starting from the microscopic formulation of the model described by
reactions, this RG flow will allow us to understand \emph{how the model parameters scale in space and time}. In turn, this information will
be helpful in determining the final equilibrium state of the aggregates of interacting species.

The sections are distributed as follows: Section (\ref{mod}) describes the models formulated and their treatments in mean field theory.
The study of the Doi-Peliti mapping and the corresponding RG flows applied to the models themselves are made ​​in sections (\ref{asexual})
and (\ref{sexual}) respectively. In section (\ref{discuss}) is made a discussion of the results. Conclusions are drawn in section
(\ref{conclusion}) and acknowledgments in section (\ref{agrada}).

\section{Models}
\label{mod}

In this section we present the two models used. They are simplified models that attempt to capture only the essential aspects of population
dynamics. The first refers to the competition between an asexual species and a pathogen that can harm
it, eventually inducing death. The second is the analogous model for the sexual species. The incorporation of the pathogen in interactions
was inspired by the Red Queen theory regarding the host's parasites. In principle, however, any other death-inducing agent can be imagined.
The models are as follows:

\subsection{Asexual species model}

The model for asexual species is described by the following reactions:

\begin{equation}
\centering
A\stackrel{\lambda}{\rightharpoonup} 2A \hspace{1.5cm} A+B\stackrel{\mu}{\rightharpoonup} B 
\label{eq1}
\end{equation}

The first reaction on the left describes the reproduction of species $A$ which occurs at rate $\lambda$ per time unit. The second
reaction describes the attack that species $ A $ suffers from its pathogen $B.$ In this attack, species $ A $ will always be
annihilated at a rate of $\mu$ per time unit per population size unit. Note that in a model that takes the spatial character of the
interactions in a $ d $ dimensional lattice, pathogen/parasites only diffuse. They are not created or annihilated. This captures their
essence of \emph{being everywhere and always seeking, by their nature, to exploit their hosts,} as mentioned above.

\subsection{Sexual species model}

The model for sexual species is described by the following reactions:

\begin{equation}
2A\stackrel{\lambda}{\rightharpoonup} 3A \hspace{1.5cm} A+B\stackrel{\mu}{\rightharpoonup} B
\label{eq2}
\end{equation}

The first reaction describes species $A's$ reproduction that occurs at rate $\lambda$ per time unit per population size unit.
Because two agents are required to reproduce a third, this reaction captures the cost of finding a mate. Everything else is as in the
previous model. 

This is not the first time that a model of this type is proposed for population dynamics incorporating the Allee effect
\cite{allee_original,allee1949principles,allee_effects} on the lattice. For recent references, see \cite{windus2007allee,gastner}. In the process known as the
quadratic contact process (QCP) \cite{durrett1999stochastic}, we also have similar reactions. QCP is sometimes called the process of sexual reproduction
\cite{preece2009sustainability}.

It is very important to keep in mind that we are considering the critical situation where the concentration of species in the lattice is initially very small
and the dynamics is dominated by diffusion \cite{mattis1998uses}. The average number of agents per site is initially much lower than 1 and sexual reproduction
is therefore penalized. Only in this critical regime does it make sense to use the renormalization group techniques.


\subsection{Mean field}
\label{mf}

In this section we consider $A$ or $B$ as the average density of agents in the sites of the lattice.

\subsubsection{Asexual model}

Using the law of mass action, we obtain the differential equations for asexual species: $\dot{A}=D_A\nabla^2A+\lambda A-\mu AB$ and
$\dot{B}=D_B\nabla^2B$ with $D_A$ and $D_B$ being diffusion coefficients. The diffusion processes only tend to homogenize populations in
space and nabla operators will be neglected from now on in this subsection. The $ B $ population is a constant on average denoted by
$\langle N_B\rangle$ and therefore $\dot{A}=(\lambda-\mu \langle N_B\rangle)A\equiv mA,$ which defines $m=\lambda-\mu\langle N_B\rangle.$ We
see that if $\mu \langle N_B\rangle < \lambda,$ $m>0$ and the $B$ population tends exponentially to infinity. Otherwise, $m<0$ and the $B$
population becomes extinct. If $m = 0,$ the $A$ population remains constant.

\subsubsection{Sexual model}

\begin{figure}[!ht]
\begin{center}
\includegraphics[width=4in]{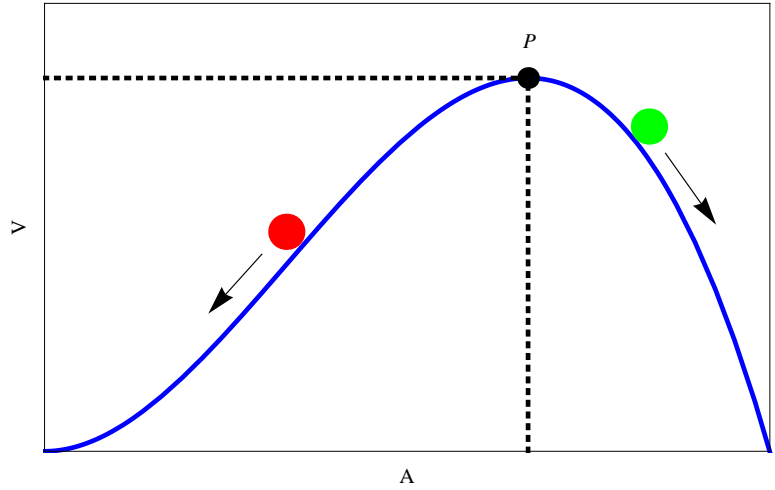}
\end{center}
\caption{
{\bf Effective potential $V(A).$} The black point $P$ has coordinates
$(A_{\textrm{max}},V_{\textrm{max}})=(\kappa/\lambda,\kappa^3/6\lambda^2).$
}
\label{V}
\end{figure}

In this case, the equation for population dynamics already disregards diffusion terms and settings $\kappa \equiv \mu\langle N_B\rangle$
is $\dot{A}=\lambda A^2-\kappa A\equiv-dV/dA$ with $V\equiv-\lambda A^3/3+\kappa A^2/2.$ $V$ is an effective potential that allows a
pictorial view of the dynamics, as illustrated in Figure (\ref{V}). The point $P$ on the potential maximum has coordinates
$(A_{\textrm{max}},V_{\textrm{max}})=(\kappa/\lambda,\kappa^3/6\lambda^2).$ For any initial population $A(0) <\kappa / \lambda,$ the 
population tends to be extinct. This fact is illustrated in figure (\ref{V}) by the tendency of the red ball to moves down the curve
to the origin. If $A(0)> \kappa / \lambda,$ population tends to infinity, a fact represented by the tendency of the green ball to
get lost in the bottomless potential hole.



In the next sections we will see how the $\kappa,$ $\lambda$ and $\mu$ parameters change with successive renormalization group iterations,
or, in other words, how the discrete nonlinear species interactions in space-time induce variations in the numerical parameter values.
These changes can transform very significantly the potential barrier to be overcome (given by $\kappa^3/6\lambda^2$) by the population.



\section{Asexual model}
\label{asexual}

In the Doi-Peliti theory, the master equation which describes the evolution of the probability vector
$|P(t)\rangle\equiv\sum_{{\cal{C}}}p({\cal{C}},t)|{\cal{C}}\rangle,$ with the sum being performed over all configurations, is written in the form
$d|P\rangle /dt = H|P\rangle ,$ where the Hamiltonian operator $H$ is composed of creation and destruction operators. Starting from this
``microscopic'' description, one derives an effective action $S$ via a path-integral representation \cite{Tauber}. Then, taking the continuum limit, one arrives
at a field theory for the model.

For the asexual model, Doi-Peliti action, already with the following Doi shifts $\tilde{\phi} \to  1+\bar{\phi},$ $\tilde{\psi}  \to
1+\bar{\psi},$ $\psi  \to  \psi+\langle N_B\rangle,$ and $\phi  \to  \phi,$ is

\begin{eqnarray}
 S[\bar{\phi},\phi,\bar{\psi},\psi]&=&\int d^dx\int
dt\Big[\bar{\phi}(\partial_t-m-D_A\nabla^2)\phi \nonumber \\
&+& \bar{\psi}(\partial_t-D_B\nabla^2)\psi-\lambda\bar{\phi}^2\phi+\mu\bar{\phi}\phi\psi \nonumber \\
&+& \mu\langle N_B\rangle\bar{\phi}\bar{\psi}\phi+\mu\bar{\phi}\bar{\psi}\phi\psi\Big]
\label{action2}
\end{eqnarray}
where $m\equiv\lambda-\mu\langle N_B\rangle$ is the bare mass. $\phi$ and $\psi$ are fields associated with the populational densities of
$A$ and $B$ respectively, while $\bar{\phi}$ and $\bar{\psi}$ are related to their statistical fluctuations. Let us assume that the
parameters $ \mu $ and $ \lambda $ are sufficiently small so that we can use the perturbation theory. Feynman diagrams associated with the
action (\ref{action2}) are shown in Figure (2).

\begin{figure}[!htb]
\begin{center}\begin{picture}(60,50)(0,-20)
  \DashArrowLine(-10,30)(-50,0){3}
  \DashArrowLine(-10,-30)(-50,0){3}
  \ArrowLine(-50,0)(-100,0)
  \Vertex(-50,0){2}
  \Text(-20,0)[]{$-\lambda\phi\bar{\phi}^2$}
  \Text(-50,20)[]{$I$}

  \ArrowLine(100,0)(70,30)
  \ArrowLine(100,0)(70,-30)
  \DashArrowLine(140,0)(100,0){3}
  \Vertex(100,0){2}
  \Text(70,0)[]{$\mu\phi\psi\bar{\phi}$}
  \Text(100,20)[]{$II$}
\end{picture} 
\end{center}

\begin{center}\begin{picture}(60,60)(0,-20)
  \DashArrowLine(-10,30)(-50,0){3}
  \DashArrowLine(-10,-30)(-50,0){3}
  \ArrowLine(-50,0)(-100,0)
  \Vertex(-50,0){2}
  \Text(-10,0)[]{$\mu\langle N_B\rangle\bar{\phi}\bar{\psi}\phi$}
  \Text(-50,20)[]{$III$}

  \ArrowLine(100,0)(70,30)
  \ArrowLine(100,0)(70,-30)
  \DashArrowLine(130,-30)(100,0){3}
  \DashArrowLine(130,30)(100,0){3}
  \Vertex(100,0){2}
  \Text(70,0)[]{$\mu\psi\bar{\psi}\phi\bar{\phi}$}
  \Text(100,20)[]{$IV$}
\end{picture}

\end{center}

\caption{Feynman diagrams}
\end{figure}
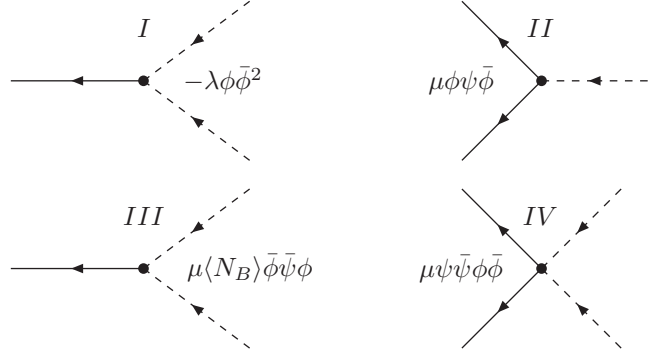

\subsection{Dynamical renormalization group}

We use the field theory techniques to find the renormalization group flow in the parameter space. The system will be analyzed using the
standard renormalization group technique, imposing the change of scale $x\to sx,$ $t\to s^zt,$ $\phi\to s^{-d-\eta}\phi,$ $\psi\to
s^{-d-\eta}\psi$ (similarly to $\bar{\phi},$ $\bar{\psi}$), and $\Lambda\to\Lambda/s,$ where $s$ is the renormalization group scale factor,
$\eta$ is a critical exponent, and $\Lambda$ is a momentum cuttoff.
Performing the standard perturbation theory procedures \cite{justin1},
using the diagrams combinations $II$ and $III$ (propagator renormalization, see Figure (3) left) and $II$ and $IV$ (vertice
renormalization, see Figure (3) right), we find the following flow equations for the model parameters in the limit of
$\Lambda\to\infty$ \cite{shnerb_discrete}:
\begin{subequations}
 \begin{align}
   \frac{d\mu}{dl}& =  \epsilon\mu+\frac{\mu^2}{2\pi \bar{D}} \label{01} \\ 
   \frac{dm}{dl}& =  2m-\frac{\langle N_B\rangle \mu^2}{2\pi\bar{D}}, \label{02}
 \end{align}
\label{flow}
\end{subequations}

\noindent where $\epsilon=2-d,$ $l=\ln{(s)},$ and $\bar{D}\equiv(D_A+D_B)/2.$ Competition parameter $\mu$ increases indefinitely with the
renormalization group iterations, favoring the extinction of the asexual species. This fact is interpreted in the last section using the
re-entrant property of diffusive systems in low dimensions.

\begin{figure}[!htb]
\begin{center}\begin{picture}(60,40)(0,-20)
  \CArc(-100,0)(15,0,80)
  \DashArrowArc(-100,0)(15,0,180){2}
  \CArc(-100,0)(15,270,0)
  \DashArrowArcn(-100,0)(15,0,-180){2}
  \Vertex(-85,0){2}
  \Vertex(-115,0){2}
  \DashArrowLine(-50,0)(-85,0){3}
  \ArrowLine(-115,0)(-150,0)


  \CArc(100,0)(15,0,80)
  \DashArrowArc(100,0)(15,0,180){2}
  \CArc(100,0)(15,270,0)
  \DashArrowArcn(100,0)(15,0,-180){2}
  \Vertex(115,0){2}
  \Vertex(85,0){2}
  \DashArrowLine(150,0)(115,0){3}
  \ArrowLine(85,0)(60,20)
  \ArrowLine(85,0)(60,-20)  

\end{picture} 
\end{center}
\caption{Propagator renormalization diagram $II+III$ (left) and vertice renormalization diagram $II+IV$ (right). }
\label{comb1}
\end{figure}
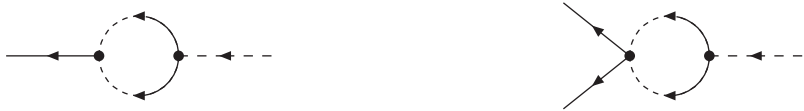

Equations (\ref{flow}) are identical to those obtained in \cite{shnerb_discrete,shnerb2001adaptation} for the reactions
$B\stackrel{\mu}{\rightharpoonup} \emptyset , A+B\stackrel{\lambda}{\rightharpoonup} 2B+A, $  with bare mass $m_{AB}\equiv\mu-\lambda\langle
N_A\rangle,$ with $\langle N_A \rangle$ representing the average number of $A.$ This model is known as the $AB$ model \cite{ABmodel}. It has
been originally proposed in \cite{shnerb_discrete,DP_shnerb2,ABmodel,competition_biology,shnerb2001adaptation} to discuss the origin of life
in terms of auto-catalysis, and it has been applied in some research areas such as ecology and economy
\cite{challet2009universal,economic_instability,yaari2008microscopic}. The spatial version of this model shows that self-replication can be
locally maintained with $B$ growing exponentially, even when average $A$ concentration would not be sufficient to sustain growth in a
homogeneous vessel. This fact is a consequence of the tendency of $\mu$ to grow with the scale $s,$ as shown by the equations
(\ref{flow}), and from the definition of $m_{AB}.$ Exactly the opposite will occur in the case of the model with asexual population, since
in this case $m=\lambda-\mu \langle N_B \rangle$ and therefore $\mu$ is subtracted rather than added.

\begin{figure}[!htb]
\centering
\subfloat[\bf RG flow for $\epsilon<0.$]{
\includegraphics[height=7cm]{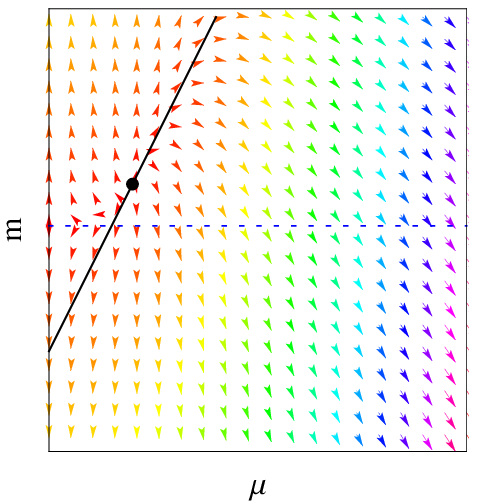}
\label{f1}
}
\quad 
\subfloat[\bf RG flow for $\epsilon\geq0.$]{
\includegraphics[height=7cm]{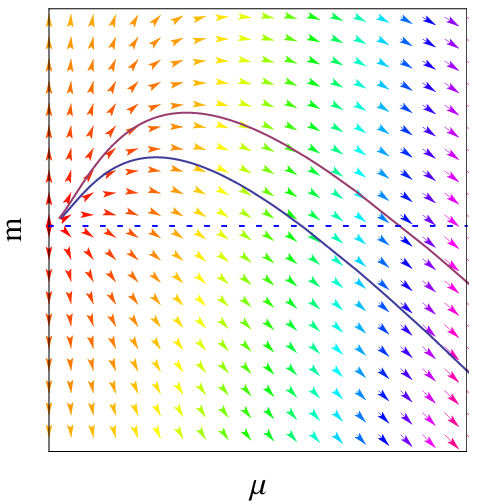}
\label{f2}
}
\caption{\bf RG flow asexual model}
\label{FF}
\end{figure}

Figures (\ref{f1}) and (\ref{f2}) show the RG flow diagrams associated with equations (\ref{flow}) for $\mu\geq0.$ On the
left we have the case of $\epsilon<0$ (or $d>2$). The black dot is the fixed point given by $(\mu^*,m^*)=(2\pi\bar{D}\epsilon,\bar{D} \pi
\epsilon ^2 \langle N_B\rangle).$ The diagonal line represents an eigenvector indicating two distinct behaviors of the diagram near the
fixed point. The horizontal dotted line represents $m = 0.$ Above the straight line and for $m> 0,$ the RG flow tends to take $m$ to
\emph{infinity}. In this case the asexual species population explodes. This happens for a sufficiently small $\mu.$ The opposite occurs
below the line (\textit{i.e.}, for sufficiently large $\mu$), with the RG flow inducing $m$ to negative values, inducing the population to extinction,
even with the mean field theory indicating explosion. We may call this phenomenon \emph{Discreteness Inducing Extinction}
(DIE).

More interesting is the figure on the right, where the DIE phenomenon is certain across the parameter space (for $\mu> 0$) for $d\leq2$
(or $\epsilon\geq0$). \emph{On the surface, asexual species always die.}




\section{Sexual model}
\label{sexual}

Let's compare the effects caused by the discrete character of the interactions in asexual and sexual reproduction. For this, we now need to
consider Doi-Peliti effective action for the sexual reproduction model:


\begin{eqnarray}
 S[\bar{\phi},\phi,\bar{\psi},\psi]& = &\int d^dx\int dt\Big[\bar{\phi}(\partial_t+\kappa-D_A\nabla^2)\phi \nonumber \\
 & + & \bar{\psi}(\partial_t-D_B\nabla^2)\psi-\lambda\bar{\phi}\phi^2+\mu\bar {\phi}\phi\psi \nonumber \\
 & + & \mu\langle
N_B\rangle\bar{\phi}\bar{\psi}\phi+\mu\bar{\phi}\bar{\psi}\phi\psi-2\lambda\bar{\phi}^2\phi^2-\lambda\bar{\phi}^3\phi^2\Big]  \nonumber \\
\label{SS}
\end{eqnarray}
where $\kappa\equiv\mu\langle N_B\rangle.$ Field interpretations are as before.

\begin{figure}[!htb]
\begin{center}\begin{picture}(60,50)(0,-20)
  \ArrowLine(80,0)(50,30)
  \ArrowLine(80,0)(50,-30)
  \DashArrowLine(120,0)(80,0){3}
  \Vertex(80,0){2}
  \Text(50,0)[]{$-\lambda\phi^2\bar{\phi}$}
  \Text(80,20)[]{$VI$}

  \ArrowLine(-20,0)(-50,30)
  \ArrowLine(-20,0)(-50,-30)
  \DashArrowLine(10,-30)(-20,0){3}
  \DashArrowLine(10,30)(-20,0){3}
  \Vertex(-20,0){2}
  \Text(-50,0)[]{$-2\lambda\bar{\phi}^2\phi^2$}
  \Text(-20,20)[]{$V$}
\end{picture} 
\end{center}
\label{novos_diagramas}
\caption{Some Feynman diagrams in sexual model.}
\end{figure}
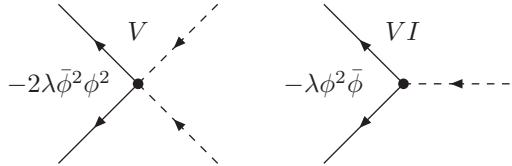

An important feature of this model is the diagram $V$ in figure (5). We also should replace the diagram $I$ in figure (2) with the diagram
$VI$ in figure (5).

\begin{figure}[!htb]
\begin{center}\begin{picture}(60,40)(0,-20)

  \CArc(0,0)(15,0,80)
  \DashArrowArc(0,0)(15,0,180){2}
  \CArc(0,0)(15,270,0)
  \DashArrowArcn(0,0)(15,0,-180){2}
  \Vertex(15,0){2}
  \Vertex(-15,0){2}
  \DashArrowLine(50,0)(15,0){3}
  \ArrowLine(-15,0)(-40,20)
  \ArrowLine(-15,0)(-40,-20)  

\end{picture} 
\end{center}
\caption{Vertice renormalization (Feynman diagram $V+VI$) for sexual model. }
\label{comb2}
\end{figure}
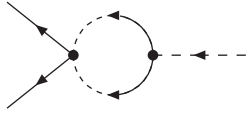





Now, the crucial point is that the parameter $\lambda$ can be explicitly renormalized using diagrams $V$ and $VI$ (see Figure (6)).
Performing the basic steps mentioned before, we have the following RG flow equations for the sexual species model:
\begin{subequations}
 \begin{align}
 \frac{d\mu}{dl}& =  \epsilon\mu+\frac{\mu^2}{2\pi \bar{D}} \label{1} \\
 \frac{d\kappa}{dl}& =  2\kappa-\frac{\langle N_B\rangle \mu^2}{2\pi\bar{D}} \label{2}  \\
 \frac{d\lambda}{dl}& =  \epsilon\lambda+\frac{\lambda^2}{\pi \bar{D}}
\label{flow_sex}
 \end{align}
\label{FLOW_SEX}
\end{subequations}

\noindent where $\epsilon=2-d,$ $l=\ln{(s)},$ and $\bar{D}\equiv(D_A+D_B)/2.$ The diagram in Figure (\ref{comb2}) which renormalizes 
$\lambda$ is equal to the diagram in Figure (\ref{comb1}) at right, which renormalizes $\mu,$ with the difference of a factor 2 in
the former. Associating this similarity to the fact that the propagators are also very similar,\footnote{The propagator for the field $\phi$
in equation (\ref{SS}) is given by $G_{\phi \phi }[k,\omega]=(D k^2+\kappa-i\tau\omega)^{-1}.$ Replacing $\kappa$ by $-m$ we obtain
the propagator for the field $\phi$ in equation (\ref{action2}). The propagators for the fields $\psi,$ $G_ {\psi\psi}[k,\omega],$ are
identical for both models.} the results are almost identical and we obtain an expression for the RG flow for $\lambda$ very similar to the
expression of $\mu.$

The new $\lambda$ RG flow does not influence $\kappa$ and therefore RG flows involving $\kappa$ and $\mu$ are as shown in figures (\ref{f1})
and (\ref{f2}) by replacing $m$ with $\kappa$ in the vertical axes. We can now examine how the potential barrier (given by
$V_{\textrm{max}}=\kappa^3/6\lambda^2$) in figure (\ref{V}) varies when the parameters are renormalized. We see that on the surface (or in
smaller dimensions: $\epsilon \geq0$), this potential barrier disappears quickly, \emph{favoring} the sexual species. This happens because
of how quickly $\lambda$ approaches infinity (with decreasing $\kappa$), making $V_ {\textrm {max}}\to0.$ The barrier that prevented the
sexual population's development is increasingly transposable if there is enough space and time for the interactions to occur. And this fact
arises from the interaction's discreteness. This is \emph{the importance of being discrete in sex}. Furthermore, according to our simplified
models, the only chance for an asexual species to exist in nature, is in a three dimensional space. This finding leads to the conjecture
that most asexual species existing today, live either in the oceans or in other ``effective'' three-dimensional media.

\section{Discussion}
\label{discuss}

A possible objection to the sexual model is that various particles can accumulate in one lattice site, leading to a divergence in the population due to the
reaction $2A\stackrel{\lambda}{\to}3A.$ In principle, the amount of particles per site can be infinite, but for our purposes, it does not matter.
One possible argument goes as follows:

If we establish the same initial conditions\footnote{Random initial positions of individuals in the lattice, without the overlapping of two or more individuals
at the same site.} for the two types of model, with both species on the verge of extinction, the sexuated species will win unequivocally, after a transient. And
this occurs twice: once because of the certainty of extinction of asexual species (at least for d = 2) and once because of the trend of sexuated species to
diverge. It does not matter if ultimately the amount of individuals per site becomes greater than one. From this point on, our minimal model is no longer valid
and constraints must be added, but the battle has already been won by the sexuated species.

Is it possible to understand physically why the description based on differential equations (or mean field theory) becomes invalid for long timeframes and large
scales in lower dimensions? To answer this question, let's first consider a different situation, for the sake of simplicity. Consider a one-dimensional lattice,
where nearly all lattice sites are populated by particles of species A, such that the only reaction occurring is the \emph{annihilation} reaction
$A+A\stackrel{\alpha}{\to}\emptyset.$ In principle, this reaction can take place everywhere. Thus, the initial dynamics is accurately described by the solution
of the kinetic rate equation $\frac{\partial\rho}{\partial t}=D_A\nabla^2\rho-\alpha\rho^2,$ where $\rho(x,t)$ is the concentration of $A$ individuals in the
lattice and $D_A$ is the diffusion constant. Therefore, initially, $\rho\propto t^{-1}.$ At later times, the lattice becomes more and more diluted and the
reaction rate is limited by the first passage time of a random walk in $d=1,$ scaling exactly with $t^{-1/2}.$ Hence, the long time behavior of $\rho$ is
renormalized to $\rho\propto t^{-1}/t^{-1/2}=t^{-1/2}.$ This is a simple example of how the reactions limited by diffusion might provide different results from
the equivalent models described in terms of differential equations.

The probability of the species $B$ (the parasites) finding the species $A$ is 1 for $d < 3$ for asexual species at the limit of the reaction limited by
diffusion because of the reentrant property of random walks in low dimensions. Physically, this means that the diffusing particle will thoroughly sweep its
neighborhood. It’s therefore is highly probable that it will react with another particle in its vicinity. Hence, it is reasonable to expect that after a short
period of time the system will be in a state where there are almost \emph{no} individuals of species $A.$ This phenomenon no longer occurs in $d > 2$ and
asexual species may exist.

In the case of sexual reproduction, there is a competition between two nonlinear phenomena: $2A \stackrel{\lambda}{\to} 3A$ and
$A+B\stackrel{\mu}{\to}\emptyset.$ As we have seen, however, the former wins from the latter and the tendency to form clusters of sexed individuals
predominates. \emph{This is due to the “shielding” effects that the boundaries of the clusters have on the individuals within it.} It is therefore reasonable to
expect that after a short period of time the system will be in a state where there are a lot of isolated particles.

\section{Conclusion}
\label{conclusion}

In this paper we propose a model that can shed some light on the question of the predominance and maintenance of sexual reproduction in nature despite all
its costs. We do not infer anything about the \emph{origin} of this predominance, but only on \emph{how} it happened. Sex is a
reproductive ritual that is inherently more complex than its rival asexual method. And this inherent complexity gives rise to some
counterintuitive features. The complexity aspect discussed here refers only to the nonlinear interactions between species and their
pathogens/parasites, and to the cost of finding a mate in the case of sexual selection. Mathematically, this cost implies a nonlinearity
(coming from the reaction $2A\to 3A$) which is absent in the asexual reproduction model. And from this nonlinearity in the interactions,
\emph{purely physical} conditions emerge that favor sexual reproduction. We need not consider anything about genetics for example.

Another important actor in this context of complexity is the discrete character of the interactions. This actor is solely responsible
for the DIE phenomenon in asexual species, where the extinction is possible for $d<2$ and certain for $d\geq2,$ even if the
mean field theory indicates otherwise. The intrinsic stochasticity induced by this discreteness is also responsible for effectively raising
the $\lambda$ parameter, as seen through the RG flow for the sexual model. This fact allows for the development of sexual populations,
despite its considerable costs for finding a mate, even in situations not covered by the mean field theory. A phenomenon able to induce this
increase, is the aggregation in clusters of interacting agents. A well-known property of diffusion is the re-entrancy of the visited sites
in low space dimensions. In particular, for $d = 1$ and $d = 2,$ the probability that the diffusing particle will ever return $(t\to\infty)$
to the starting point is equal to $1.$ Physically, it means that the diffusing particle thoroughly sweep its neighborhood and thus
it is highly probable that it will react with another particle in its vicinity. Hence, it is reasonable to expect that after short period of
time the system will be in a state where there is a lot of isolated particles.\footnote{A similar phenomenon also occurs among species and
their pathogens/parasites, making the death rate also rises, explaining the extinction of asexual species in $d \leq 2.$ However,
this increase is insufficient to overcome the increased of the sexual species for $d\leq2.$} The clustering of sexual agents favors a
localized increase in the $\lambda$ reproduction rate, allowing their permanence and development. We must not forget that an individual
cannot reproduce arbitrarily fast. This imposes an upper limit for $\lambda.$ Briefly, I propose that: 

\begin{center}
\emph{The explanation for the maintenance of sexual reproduction in nature is in the scale of populations, far above the molecular scale of the gene, and
manifests itself as an \textbf{emergent property} of the discrete interactions in the intermediate scale of the individuals.}
\end{center}

\section{Acknowledgments}
\label{agrada}

RVS is grateful to Ronald Dickman and Linaena M\'ericy da Silva for helpful comments. This work was supported by the Conselho Nacional de
Desenvolvimento Cient\'ifico e Tecnol\'ogico, Brazil.

\bibliographystyle{plos2009.bst}
\bibliography{bibliografia.bib}

\end{document}